\documentclass[12pt,draft]{article}

\usepackage[english]{babel}

\usepackage{amssymb}
\usepackage{latexsym}
\usepackage[round]{natbib}
\usepackage{url}
\usepackage{multirow}

\usepackage[final]{graphicx}

\usepackage{float}
\restylefloat{table}

\usepackage{color}

\renewcommand{\Re}{\mathbb{R}}

\newcommand{\prob}{\mathrm{Pr}}
\newcommand{\vat}{\mathbb{E}}

\newcommand{\g}{\,\vert\,}
\newcommand{\bc}{\begin{center}}
\newcommand{\ec}{\end{center}}

\newcommand{\bitem}{\begin{itemize}}
\newcommand{\eitem}{\end{itemize}}

\newcommand{\be}{\begin{eqnarray*}}
\newcommand{\ee}{\end{eqnarray*}}
\newcommand{\ben}{\begin{eqnarray}}
\newcommand{\een}{\end{eqnarray}}

\usepackage{layout}

\renewcommand{\baselinestretch}{1.5}

\begin{document}

\title{Objective Bayesian Comparison of Constrained Analysis of Variance Models }
\author{
Guido Consonni\\
Universit\`{a} Cattolica del Sacro Cuore, Milan, Italy\\
\url{guido.consonni@unicatt.it}
\and
Roberta Paroli \\
Universit\`{a} Cattolica del Sacro Cuore, Milan, Italy\\
\url{roberta.paroli@unicatt.it}
}
\date{  }
\maketitle


\begin{abstract}
In the social sciences we are often interested in comparing models
specified by parametric equality or inequality constraints. For
instance, when examining three group means $\{ \mu_1, \mu_2,
\mu_3\}$ through an analysis of variance (ANOVA),  a model may
specify that $\mu_1<\mu_2<\mu_3$, while another one may state that
$\{  \mu_1=\mu_3\} <\mu_2$, and finally  a third model may instead
suggest that  all means are unrestricted.
This is a challenging
problem,  because it involves a combination of non-nested models,
as well as nested models having the same dimension.
We adopt an objective  Bayesian approach, 
and derive the posterior
probability of each model under consideration.
Our
method is based on  the intrinsic prior
methodology, with suitably modifications to accommodate equality and inequality  constraints.
Focussing on normal ANOVA models, a comparative
assessment
is carried out through
simulation studies, showing that correct model identification is possible even in situations where frequentist power
is low.  We also present an application to real data
collected in a  psychological experiment.
\end{abstract}


\section{Introduction}

In this paper we consider the comparison of models specified by
inequality or equality constraints on its parameters, or possibly
by a combination of them. These models are common in the social
sciences; see \cite{Klugkistetal:2005} and \cite{VanWesel:2011}.
For instance,  consider a three-way normal ANOVA with group
means $\mu_j$. One possible model is $M_1:$ $\mu_1<\mu_2<\mu_3$,
while another one is $M_2:$
 $\{ \mu_1 = \mu_3 \} < \mu_2$.
 Two special models stand out: the unconstrained, or encompassing, model $M_e:$ $ \mu_1,  \mu_2, \mu_3$, wherein no constraint is imposed on the parameters, and the null model
 $M_0:$ $ \mu_1 = \mu_2 =\mu_3$.

 Consider two sampling  models for the
observables $y$, namely
$M_1: \{ f_1(y|\theta_1 \in \Theta_1)\}$ and
$ M_2: \{ f_2(y|\theta_2 \in \Theta_2)\}$.
Let $p_1(\theta_1)$ and $p_2(\theta_2)$ be the priors on the parameters under each of the two models.
The Bayes Factor (BF) of model $M_1$ against model $M_2$ for given data $y$ is the ratio $m_1(y)/m_2(y)$ of the two marginal likelihoods, where $m_i(y)=\int _{\Theta_i} f_i(y|\theta_i)p_i(\theta_i) d\theta_i$.
If prior model probabilities are added,  then one can also compute the posterior model probabilities.
Of particular interest is the case in which $M_1$ is nested into $M_2$, that is the two sampling densities belong to the same family with $\Theta_1   \subset \Theta_2$.
  In this paper  we consider  general nested situations where the two parameter subsets may have different dimensions, as  in the ANOVA models $M_2$ and $M_e$ above, as well as  the same dimension, as  in the case of $M_1$ and $M_e$.


We follow  an objective Bayesian standpoint; see \cite{Berg:2006}.
%
Specifically, the focus of this work is objective Bayesian model  selection, which has led to specific techniques for the
construction of prior distributions, quite separate from estimation:
intrinsic prior
\citep{Berg:Peri:1996,More:1997};
fractional Bayes factor \citep{Ohag:1995}; expected posterior
prior \citep{Pere:Berg:2002}; a comprehensive  review is in
\citet{Pericchi:2005}.
In particular, the intrinsic prior approach, and its generalization based on the expected posterior prior,
have emerged as a powerful methodology for comparing nested models in a variety of settings; see for instance
%
\cite{Case:More:2006}, \cite{GiroEtAl:2006},
\cite{Cons:LaRo:2008},
\cite{Leon-NoveloEtAl:2012}. \citet{Case:EtAl:2009} and \citet{More:Etal:2010}
deal with consistency issue.

Far less attention has been devoted to the Bayesian comparison of constrained
models specified by inequality/equality constraints; let alone its objective counterpart. Early stylized analyses appeared in
  \cite{CanoEtAl:2008} and \citet{Moreno:2005}, essentially   dealing
with one sided hypothesis testing.
 \cite{Klugkistetal:2005}, \cite{Klug:Hoij:2007},
\cite{LaudyEtAl:2007}  have introduced a methodology, named
\emph{encompassing} prior, which deals specifically with
inequality constrained models. For a critical discussion, see \citet{Stern:2005}.
Objective Bayesian methods for the comparison of inequality constrained models are presented in \citet{Hoij:2013} for general models and in  \citet{VanWesel:2011} for ANOVA models.
The latter work
contains some critical features which   our approach tries to overcome. In particular, we develop an alternative fully automatic procedure, which does not require parametric fine-tuning,    nor empirical training samples,  and
can deal simultaneously with inequality and
equality constraints (the latter being treated \emph{exactly}).

The rest of the paper is organized as follows. Section \ref{sec:Bayesian comparison} deals with conceptual issues related to the Bayesian comparison of nested models; section \ref{sec:Objective priors}  presents the general framework of our methodology, which is implemented for ANOVA models in section \ref{sec:Intrinsic}. Section \ref{sec:Applications} presents simulations and an application. Finally, section \ref{sec:Discussion} contains a brief discussion.

\section{Bayesian comparison of nested   models}
\label{sec:Bayesian comparison}

Consider for example the comparison of the ANOVA normal model $M_0:$ $\Theta_0=\{ (\mu_1,\mu_2,\mu_3) \in \Re^3:\mu_1=\mu_2=\mu_3  \}$
against the unrestricted model $M$: $\Theta=\{ (\mu_1,\mu_2,\mu_3) \in \Re^3  \}$ (for simplicity we equate  the parameter space with the space of means and omit nuisance parameters). In a Bayesian setting  this would usually proceed by designating $\mu_0 \in \Re$ as the unique parameter indexing model $M_0$, and assigning a prior $p_0(\mu_0)$; similarly   a prior $p(\mu)$ would be assigned to $\mu^T=(\mu_1,\mu_2,\mu_3)$. The two priors would lead to the corresponding marginal distribution evaluated at the data $y$, $m_0(y)$ and $m(y)$,  and then to $BF_{M_0,M}(y)=m_0(y)/m(y)$.
Usually $p(\mu)$ is continuous, and therefore assigns probability zero to    the set $\Theta_0$; accordingly one could redefine $M$ as $M^{\prime}$ having parameter space $\Theta^{\prime}=\Theta \setminus \Theta_0$ and the resulting marginal distribution and BF would be identical.
The bonus of this fact is that now $M_0$ and $M^{\prime}$ are distinct and so the comparison is meaningful; in particular we can assign positive prior probabilities to   $M_0$ and $M^{\prime}$ summing to one.
Notice that the posterior probability of  $M_0$ may well exceed the posterior probability of $M^{\prime}$ (for instance this will occur if $BF_{M_0,M^{\prime}}>1$ and the probability mass is uniformly distributed  on the two models).
This happens because the BF exhibits a natural Occam's razor which is incorporated into the marginal distribution $m_0(y)$.
The  procedure described so far is routinely carried out in Bayesian model comparison;  see for instance \cite{Lian:Paul:etal:2008} in the context of  Bayesian  variable selection.




         Now consider the comparison of two nested models whose parameter spaces  have the same dimension.
For concreteness let $M_1:$ $\mu_1<\mu_2<\mu_3$,  while $M$ still  denotes the unrestricted model. Let $\Theta_1$ represent the parameter space under $M_1$.
 Since  $\Theta_1 \subset \Theta$, it follows that $M_1$ implies $M$. Additionally, $\prob(\Theta_1 \g M)>0$;   hence  the two models cannot be made distinct and Bayesian model comparison is ill-posed.
The above argument  extends to a collection of subsets $\{ \Theta_i \subset \Theta \}$ having positive probability under $M$.
We can compute the probability  of the sets $\Theta_i$ under $M$.
This is fine if parametric inference is the goal; in particular,  posterior probabilities may be useful for exploratory analysis by pointing to regions of the parameter space which are  supported by the data, and that we may have not considered \emph{a priori} likely; see \citet{Stern:2005}.
However, if subsets $\Theta_i$'s represent a collection of scientific theories (models),  probabilities of sets are not satisfactory for comparison purposes, because they fail
to incorporate a penalty for complexity: trivially,  the larger the size of $\Theta_i$, the higher its probability; see on this issue \citet{Klugkistetal:2005}.

The natural way out of this  difficulty is to realize that when using the BF we are actually comparing Bayesian models, as opposed to sampling models (or subsets).
Let  model $M_i$ correspond to an arbitrary  subset $\Theta_i \subset \Theta$, $\mbox{dim}(\Theta_i) \leq \mbox{dim}(\Theta)$; let $p(\theta \g M_i)$ be the parameter prior under $M_i$. The Bayesian model is the pair $B_{M_i}=\{  M_i, p(\theta \g M_i) \}$.
The two models $B_{M_i}$ and  $B_{M}=\{  M, p(\theta \g M) \}$ are two distinct data-generating mechanisms (through their marginal distributions). In particular we can no longer state that if $B_{M_i}$ holds, so does  $B_{M}$.
We can thus freely
     assign prior probabilities on the space of  Bayesian models.   For instance   $\prob(B_{M})<\prob( B_{M_i})$ is permissibile, if the latter  corresponds to a scientific theory which is believed to be highly reasonable, at least \textit{a priori}.

 Of course, from a substantive viewpoint,  it is important to keep track of the origin of each Bayesian model (sampling model and parameter prior): in particular, if $M_i$
  is nested into $M$ through a parametric constraint, then it still makes sense to refer to $B_{M_i}$ as a constrained model, because  this feature has been  incorporated into the marginal distribution of the observables, by integrating  the sampling density  with respect to a prior having smaller support  than under the unconstrained model. In this way,  the parsimony of $M_i$  extends to $B_{M_i}$,  thus enforcing a natural Occam's razor; see subsection \ref{subsec:Encompassing} for an illustration.
  This parallels what happens   when the nested model $M_i$ is of lower dimension than $M$, so that the marginal distribution is an expectation over a lower-dimensional parameter space.

In conclusion, there is  no  conceptual distinction between the Bayesian comparison of nested models when the  dimensions of the two models are equal or different, if one relies on the notion of a Bayesian model.
Once this concept is understood, we may still use the notation $M_i$ even when we refer to the Bayesian model for comparison purposes. We will follow this convention in the rest of this paper.

\section{Objective Bayesian comparison of constrained models}
\label{sec:Objective priors} 
In this section we develop a general methodology for the objective Bayesian comparison of  models with inequality/equality constraints,  which will be applied to ANOVA models in section \ref{sec:Intrinsic}.

\subsection{Intrinsic priors}

Consider two sampling models for the observables $y$, namely
$M_1: \{ f_1(y|\theta_1)\}$ and
$ M_2: \{ f_2(y|\theta_2)\}$.
Let $p_1^N(\theta_1)$ and $p_2^N(\theta_2)$ be estimation based priors (e.g.
 reference priors, or other conventional priors; here the superscript \lq \lq $N$\rq \rq{} stands for noninformative).
There are two reasons why such priors are not suitable for
testing or model choice: i) they are typically improper; ii) each prior is
exclusively based on its own model, and thus the two priors are
not \lq \lq linked\rq \rq{}. Point i) implies the well acknowledged fact that the BF is  defined only up to an arbitrary constant.
Point ii) is less  known, but equally crucial, and is related to compatibility of priors across models;  see \citet{Cons:Vero:2008} for some general discussion,
and \citet{Cons:LaRo:2008} and \citet{Case:More:2009} with specific reference to intrinsic priors.

To deal with i) partial BFs were first introduced followed by a
more robust version, namely   intrinsic BF (IBF); see
\cite{Berg:Peri:1996}. The IBF is asymptotically equivalent to a
an actual BF computed using a pair of \emph{intrinsic} priors (one
under each model).
If $M_1$ is nested  into, and of lower dimension than,
$M_2$, the intrinsic prior for $\theta_1$ coincides with the original
prior, i.e. $p_1^I(\theta_1)=p_1^N(\theta_1)$. On the other hand,  the
intrinsic prior for $\theta_2$ can be constructed in two steps
\begin{itemize}
\item[i)] Conditional intrinsic prior (CIP)
\ben
\label{CIP}
p_2^I(\theta_2|\theta_1)=p_2^N(\theta_2) \vat^{M_2}_{\theta_2}
\left(  \frac{f_1(x|\theta_1)}{m_2^N(x)} \right)
\een
\item [ii)] Intrinsic prior (IP)
\be
p_2^I(\theta_2)= \int p_2^I(\theta_2|\theta_1) p_1^N(\theta_1) d\theta_1,
\ee
\end{itemize}
where $m_2^N(x)=\int f_2(x |\theta_2)p_2^N(\theta_2) d\theta_2$, and the expectation appearing in (\ref{CIP}) is with
respect to the sampling distribution under $M_2$, $f_2(x|\theta_2)$, where $x$ is
a random vector of  minimal sample size (so that $0<m_2^N(x)<\infty$, for all $x$). It can be verified that the
CIP  $p_2^I(\theta_2|\theta_1)$  is always
proper,
while the intrinsic  prior  $p_2^I(\theta_2)$ may
be improper. Clearly neither CIP nor IP  depend on data.

 A CIP is  tailored to the comparison  of model
$M_2$ relative to  $M_1$. In particular,  $p_2^I(\theta_2|\theta_1)$ accumulates more mass
 than $p_2^N(\theta_2)$  around the parameter  subspace which characterizes
$M_1$. This is a very reasonable property,
because  it makes the comparison of the two models fairer,
especially in the most critical situation, that is when the data tend
to support the smaller model $M_1$; for further discussion on this point see
\citet{Cons:LaRo:2008} and \citet{Cons:More:Vent:2011}. A similar
property is of course enjoyed by the intrinsic prior $p_2^I(\theta_2)$,
because it is an average (possibly with respect to an improper
measure) of conditional intrinsic priors.
CIP and IP are an effective way
of  \lq \lq linking\rq \rq{} the priors under the two models being compared.

The BF of model $M_2$ against $M_1$  under the CIP is given by
 \ben \label{BFCIP} BF_{21}^{IP}(y|\theta_1)=\frac{\int
f_2(y|\theta_2)p_2^I(\theta_2|\theta_1)d\theta_2}{\int
f_1(y|\theta_1)p_1^N(\theta_1)d\theta_1}; \een
A similar calculation could be done under the IP, but is omitted because it will not be used in this paper.
Clearly, since $\theta_1$ is unknown, $BF_{21}^{IP}(y|\theta_1)$
is of no direct use; however we like to single it out, because it
will play a special role in our method.

%

The above procedure is useful also for comparing  two non-nested models, $M_1$ and $M_2$, if one can identify a model $M_0$
 which is nested in both $M_1$ and  $M_2$, and is of lower dimension that either model.  In this way the
comparison within the two pairs $\{ M_1,M_0 \}$ and $\{ M_2,M_0
\}$ can be carried out through $BF_{10}^{CIP}(y)$ and
$BF_{20}^{CIP}(y)$,  from which $BF_{21}^{CIP}(y)=BF_{20}^{CIP}(y)
\times BF_{01}^{CIP}(y)$ can be coherently deduced,  because
$m_0(y)$, the marginal distribution of the data under $M_0$, is
the same under the two distinct BFs.

 \subsection{Encompassing and truncated priors}
 \label{subsec:Encompassing}

Consider a  model  $M_e$, and let $\theta \in \Theta$ be its parameter.
 We assume that
$\Theta$ is an unrestricted Euclidean space of the appropriate dimension.
Define a collection of \textit{constrained} models $\{ M_k \}$. Let $\Theta_k \subset \Theta$ denote the constrained
parameter subset characterizing $M_k$. Since $\Theta$  contains each $\Theta_k$,
we refer to $M_e$ as the \textit{encompassing} model.

A natural way to compare the models $\{ M_k \}$ is to assign a unique \emph{proper} prior  to
$\theta$  under $M_e$, $p(\theta|M_e)$, having support $\Theta$.
Next,
assuming for the moment only
 \emph{inequality} constraints,
the parameter prior under
$M_k$, $p(\theta|M_k)$,  can be \emph{derived} by truncating
$p(\theta|M_e)$
to the subspace $\Theta_k \subset \Theta $.
Since $M_k$ is defined only through inequality constraints,
$\dim(\Theta_k)=\dim(\Theta)$; accordingly we still denote  with
$\theta$ the parameter for model $M_k$, and append the model
symbol as a conditioning event in the prior.
%
This top-down  assignment across parameter spaces, also called encompassing-prior approach,  establishes a natural link between all priors.

Consider now the BF of model $M_k$, equipped with its restriction
prior $p(\theta|M_k)$,
\emph{versus} the encompassing model $M_e$, with prior $p(\theta|M_e)$. It can be checked that
\ben \label{BFEncompassing}
BF_{ke}(y)=\frac{\prob(\theta \in \Theta_k|y,M_e)}{\prob(\theta \in \Theta_k|M_e)};
\een
see also \citet{Klug:Hoij:2007}. The quantity in (\ref{BFEncompassing}) is  the relative belief ratio of
subset $\Theta_k$, as described in \citet{Bask:Evan:2013}, and
  is related to the Savage-Dickey density; see \citet{WetzelsEtAl:2010}.
Notice the simplicity of this calculation, and how  it  automatically
adjusts for model complexity. In particular, if   $\Theta_k$ is
very "small" relative to $\Theta$, then both the numerator and denominator of (\ref{BFEncompassing}) are
also likely to be very small; yet $BF_{ke}(y)$ can be very high.

%


The encompassing/truncation approach was presented assuming that the various
submodels had been specified exclusively by inequality
constraints. The reason is that strict positivity of the numerator and denominator of
(\ref{BFEncompassing})  breaks
down if $\Theta_k$ is specified also by means of  equality
constraints,  under standard continuous priors
$p(\theta|M_e)$. To solve this difficulty,  \citet{Klug:Hoij:2007}
advocate using \emph{about equality} constraints. This is
equivalent to approximating a point hypothesis $\theta=\theta_0$
through an interval hypothesis $|\theta-\theta_0|<b$.
Besides being \emph{ad hoc},  this method raises the usual
question of how to fix $b$. \citet{VanWesel:2011} develop a method
to compute the BF of an  equality constrained model against the
encompassing model through a sequence of \lq \lq about equality
constrained models\rq \rq{} corresponding to a decreasing sequence
$\{ b_r \}$, $r=1,\ldots,R$, $R\rightarrow\infty$,  until
stabilization in the result takes place.

\subsection{Bayes factors and posterior model probabilities}
\label{subsec:Intrinsic} 

In this subsection we present a novel
proposal for constructing objective priors for comparing
constrained models, where the constraints can involve  inequalities, as well as
equalities,  among the components of the parameter
vector $\theta \in \Theta$.

%
Consider a general constrained model $M_k$, possibly involving both equality and inequality constraints, and  characterized by a
parameter subspace $\Theta_k \subset \Theta$.
Define the \emph{encompassing}-$M_k$
model, written $M_{e(k)}$,  as that model whose parameter space
$\Theta_{e(k)}$ has the same \emph{equality} constraints as
$\Theta_k$, whereas the \emph{inequality} constraints are entirely
relaxed.  Notice that dim($\Theta_{e(k)}$)=dim($\Theta_k$). In
particular $M_{e(k)}$ may coincide with a model in the list of
entertained models, or it may be a new, additional  model; section \ref{sec:Intrinsic} will illustrate this point.


We also introduce the null sampling  model
$
 M_0: \{ f(y|\theta_0, M_0), \,  \theta_0 \in \Theta_0   \},
 $
with the requirement that it be nested in all the \emph{encompassing}-$M_k$
models under consideration.
%

 We now turn to  the specifications of prior distributions.
 \begin{itemize}
 \item
 For each model $M_k$ identify the corresponding $M_k$-encompassing model $M_{e(k)}$. Let $p^N(\theta_{e(k)}|M_{e(k)})$ be its default prior.

 \item
 Compute the conditional intrinsic prior for $\theta_{e(k)}$, given $\theta_0$, under model $M_{e(k)}$
 \ben
 \label{CIP4Me(k)}
 p^I(\theta_{e(k)}|\theta_0, M_{e(k)}),
 \een
as in (\ref{CIP}). Recall that $p^I(\theta_{e(k)}|\theta_0, M_{e(k)})$ is a  proper distribution.
 \item
 Define the parameter prior under $M_k$, conditional on $\theta_0$, by restricting  (\ref{CIP4Me(k)}) 
 to the subspace $\Theta_k$
 \ben
 \label{CIPMk}
 p(\theta_{k}|\theta_0, M_k)=
 c_k(\theta_0)p^I(\theta_{e(k)}|\theta_0, M_{e(k)})1(\Theta_k),
 \een
 where
$1/c_k(\theta_0)=\int_{\Theta_k}p^I(\theta_{e(k)}|\theta_0, M_{e(k)})
 d\theta_{e(k)} $.
 Recall that $0<c_k(\theta_0)<\infty$ because $p^I(\theta_{e(k)}|\theta_0)$ is proper.
 \end{itemize}

We are now ready
to compute the BF for $M_k$, relative to $M_0$ for every submodel $M_k$, and combining the results as in (\ref{probIP}).
 \begin{itemize}
 \item
 Using the encompassing prior approach, we first compute
 \ben
 \label{BFCIPke(k)}
 BF^{IP}_{k,e(k)}(y|\theta_0)=c_k(\theta_0)/d_k(y,\theta_0),
 \een
 as in equation (\ref{BFEncompassing}) where $c_k(\theta_0)$ is defined in equation (\ref{CIPMk}), and
 $1/d_k(y,\theta_0)=\int_{\Theta_k}p^I(\theta_{e(k)}|y, \theta_0, M_{e(k)})d\theta_{e(k)}$.
 \item

Using the standard intrinsic prior approach for nested models, we compute the BF based on the
conditional intrinsic prior  $BF_{e(k),0}^{IP}(y|\theta_0)$. This
can be done  as in  (\ref{BFCIP}) replacing  $\theta_2$ with $\theta_{e(k)}$, and $\theta_1$ with $\theta_0$.

\item
Finally
\ben
\label{BFk0}
BF_{k0}^{IP}(y|\theta_0)=BF_{k,e(k)}^{IP}(y|\theta_0) \times
BF_{e(k),0}^{IP}(y|\theta_0).
 \een
Notice that $BF_{k0}^{IP}(y|\theta_0)$
is well defined, because the prior for
$\theta_{e(k)}$ under model $M_{e(k)}$ is the same in the two BFs
appearing on the right-hand-side, namely
$p^I(\theta_{e(k)}|\theta_0)$, and the same applies to the
marginal  densities for $y$ under $M_{e(k)}$, which therefore
cancel out.

 \end{itemize}

Having obtained $BF_{k0}^{IP}(y|\theta_0)$ as in (\ref{BFk0}) the
posterior probability of $M_k$, given $\theta_0$ is readily
available \emph{via} the formula
\ben
\label{probIP}
\prob^{IP}(M_k|y,\theta_0)=\left( 1+ \sum_{l \neq k}
\frac{p_l}{p_k} BF^{IP}_{lk}(y|\theta_0) \right)^{-1},
\een
where
$BF^{IP}_{lk}(y|\theta_0)=BF^{IP}_{l0}(y|\theta_0) \times
BF^{IP}_{0k}(y|\theta_0)$ with $BF^{IP}_{l0}(y|\theta_0)$
calculated as in (\ref{BFk0}), and $p_k=\prob(M_k)$ is the prior
probability of model $M_k$.

All the calculations performed so far are conditionally on
$\theta_0$.
Eventually, in order to implement our procedure, a value for
$\theta_0$ has to be fixed.
Note that the dimension of $\theta_0$
is typically  very low, because it indexes  the null model.
For instance,   in the
normal ANOVA model considered in the next section
$\theta_0$ is a two-dimensional vector.
 Accordingly, one
can estimate $\theta_0$ very efficiently under $M_0$,
possibly using a maximum likelihood estimate $\hat{\theta}_0$, or a Bayesian estimate based on a noninformative prior $p^N(\theta_0 |M_0)$.

%
%


\section{ANOVA models}
\label{sec:Intrinsic} 

In this section we provide a detailed
analysis of constrained normal ANOVA models using the methodology
described in subsection \ref{subsec:Intrinsic}.
The encompassing sampling model for the observables, conditional on
$(\mu_1, \ldots, \mu_J)$ and $\sigma$, is \be y_{ij}=\mu_j
+\epsilon_i, \ee
where $i=1, \ldots, n$ denotes units,  $j=1, \ldots, J$ groups.
The mean structure is unconstrained so that $(\mu_1, \ldots,
\mu_J) \in \Re ^J$, while the error term satisfies the usual
assumption of linear regression models, namely $\epsilon_i|\sigma
\stackrel{iid} \sim N(0, \sigma^2)$.
For concreteness and motivation,  we start by considering  an ANOVA  model choice setting presented in  \citet{Lucas:2003}, and further analyzed in \citet{VanWesel:2011}; in this way  we can illustrate the implementation of our method directly on this problem.    
The data originate from a psychological experiment measuring the attitude of subjects classified in $J=5$ groups; see also subsection \ref{subsec:lucas}.
Four models of interest (theories) are identified in terms of
relationships among the group means  \be
M_e&:&  \mu_1, \mu_2, \mu_3, \mu_4, \mu_5 \\
M_a&:&  \mu_2< \mu_1 <\mu_4< \{\mu_3= \mu_5 \} \\
M_b&:&  \{\mu_1, \mu_3 \} > \{\mu_2, \mu_4, \mu_5 \} \\
M_0&:&  \mu_1=\mu_2= \mu_3= \mu_4= \mu_5,
\ee
where $M_e$ is the encompassing model, $M_0$ the null model, and $M_a$ and $M_b$ are intermediate models.
 Let $y^T=(y_{11}, \ldots, y_{1n_1}, \ldots y_{J1}, \ldots,
y_{Jn_J})$ denote the vector of responses.
 Under the usual normal set-up, we can rewrite model $M_0$ as
\be
M_0: \, y= \alpha_0 1_n+ \epsilon_0,
\ee
where  $1_n$ is an $n=(n_1+\ldots,n_J)$-dimensional vector
with all components equal to 1, $\alpha_0$ is the common mean,   and
$\epsilon_0|\sigma_0 \sim N_n(0_n,\sigma_0^2 I_n)$, where $0_n$ is the $n$-dimensional vector with all components equal to 0, and  $I_n$ is the identity matrix of order $n$. On the other hand
$M_e$ can be written as
\ben \label{MfsamplingModel} M_e: \, y=  \alpha 1_n +X
\delta + \epsilon,
\een
where
$\alpha$ is the mean of group 1, $\delta^T=(\delta_2,
\ldots, \delta_J)$ represents the additional mean effect of group
$j=2, \ldots, J$, relative to  group 1.
$X$ is a $n \times (J-1) $  matrix,
with column $j$   containing a one in positions
corresponding to units  in group $j$ ($j=2, \ldots, J$), and zeros
otherwise.
Finally,
$\epsilon|\sigma \sim N_n( 0_n ,\sigma^2 I_n)$.

The  mean structure of the constrained model $M_a$
can be rewritten as \be M_a:
\delta_2<0,   0<\delta_{3}> \{ \delta_{2}, \delta_4 \},  \delta_4>0,
\ee
where $\delta_3$ is the additive term that appears for units in
group $j=3$ and $j=5$.
We used the convention that, whenever   an equality
constraint is established among a subset of group-means, the
corresponding $\delta$ is indexed by the lowest index  of the
original constituent groups.

On the other hand, the mean structure of the constrained model $M_b$ can be rewritten as
\be
M_b:
\{ \delta_2, \delta_4, \delta_5 \} <0,  \delta_{3}> \{ \delta_2,
\delta_4, \delta_5 \}; \quad \delta_3=\delta_5. \ee
Finally, the
encompassing-$M_a$ model, $M_{e(a)}$,  can be written as \be
M_{e(a)}: y=\alpha_{e(a)} 1_n +X_{e(a)}\delta_{e(a)}+\epsilon_{e(a)}, \ee
where
$\delta_{e(a)}^T=(\delta_{e(a),2}, \delta_{e(a),3},
\delta_{e(a),4})$ is a three-dimensional vector whose components
represent, in the order, the mean excess (relative to group 1) of
group 2, group 3=group 5, and of group 4.
We
emphasize that $M_{e(a)}$ is an encompassing model because its
parameters are free to vary without constraints.

 \vspace{0.6 cm}
 We now return to a general formulation.
Consider a constrained ANOVA model $M_c$. Let $M_{e(c)}$ be the
corresponding encompassing-$M_c$ model, and denote its parameter space by
$\Theta_{e(c)}$. Let $\Theta_c \subset \Theta_{e(c)} $ be the
parameter space of $M_c$ characterizing its \emph{inequality}
constraints relative to $M_{e(c)}$ (recall that $M_c$ is
distinguishable from $M_{e(c)}$ only by means of inequality
constraints). The goal is to compute the BF of $M_c$ against the
null model $M_0$ based on the conditional  intrinsic prior
procedure described in subsection \ref{subsec:Intrinsic}. This will be achieved in three steps.
i)
$BF^{IP}_{M_{e(c)},M_0}(y|\alpha_0,\sigma_0)$;
ii)
$BF^{IP}_{M_c,M_{e(c)}}(y|\alpha_0,\sigma_0)=
\frac{\prob^{IP} \{ \theta \in \Theta_c|y,\alpha_0, \sigma_0, M_{e(c)}  \}}
{\prob^{IP} \{ \theta \in \Theta_c|\alpha_0, \sigma_0, M_{e(c)}  \}}$;
iii) finally
\ben
\label{BFIPMcM0}
BF^{IP}_{M_c,M_0}(y|\alpha_0,\sigma_0)=
BF^{IP}_{M_{e(c)}, M_0}(y|\alpha_0,\sigma_0) \times
BF^{IP}_{M_c,M_{e(c)}}(y|\alpha_0,\sigma_0).
\een

%
%

We will  examine the first two  steps separately below.

\subsection{Bayes factor of an encompassing model relative to the null model}

Consider a constrained ANOVA model $M_c$ and its
encompassing-$M_c$ model $M_{e(c)}$. With slight abuse of
terminology,  we name the latter $M$ to simplify notation. Clearly
$M$ contains only unconstrained parameters, and we write its
Bayesian version as
 \ben
 \label{MsamplingModel}
 \{ f(y|\alpha, \delta, \sigma, M)=
N_n(y|\alpha 1_n +X \delta, \sigma^2 I_n)
, \, p^N(\alpha,\delta,\sigma|M) \propto 1/\sigma \}.
\een
Consider the comparison of the pair $(M,M_0)$ and the
corresponding intrinsic priors.
We have
\ben \label{CIPalphadeltasigma} p^I( \alpha, \delta,
\sigma|\alpha_0, \sigma_0, M)= \frac{2}{\pi \sigma_0
(1+\frac{\sigma^2}{\sigma_0^2})} N_J(\alpha_0 e,
(\sigma^2+\sigma_0^2)W^{-1}), \, \alpha \in \Re, \delta \in
\Re^{J-1},  \sigma>0, \een
where $e^T=(1, 0, \ldots, 0)$, $W^{-1}=\frac{n}{J+1}(Z^{T}Z)^{-1}$, with $Z  \equiv(1_n \vdots X)$.
%


Result (\ref{CIPalphadeltasigma}) is  standard  in the intrinsic prior methodology for normal linear regression models; see for instance
\citet[formula (4)]{GiroEtAl:2006},  and references therein.
Recall that the conditional intrinsic prior (\ref{CIPalphadeltasigma}) is proper. Moreover
\ben
(\alpha,\delta|\sigma, \alpha_0, \sigma_0, M) &\sim& N_{J}(\alpha_0 e, (\sigma^2+\sigma_0^2)W^{-1}) \label{CIPalphadeltasigma1}\\
(\sigma^2|\sigma_0, M) &\sim& InvBeta(\frac{1}{2},\frac{1}{2},\sigma_0^2)
, \label{CIPalphadeltasigma2}
\een
where $InvBeta(a,b,c)$, $a>0, b>0, c>0$, is an inverted-beta density with parameters $(a,b,c)$ having density
\be p(v|a,b,c)=\frac{c^b}{B(a,b)}v^{a-1} \left(\frac {1}{v+c}
\right)^{a+b}, \quad v>0, \ee  \citep[p. 221]{Raiffa:Schl:1961}.
%
%

Since
$BF^{IP}_{M,M_0}(y|\alpha_0,\sigma_0)=\frac{m^{I}(y|\alpha_0,\sigma_0,
M)}{f(y|\alpha_0,\sigma_0,M_0)}$, we only require the computation
of the numerator, because the denominator is immediately available
and is given by $N_n(y|\alpha_0 1_n, \sigma_0^2 I_n)$.

Consider now \ben \label{mIP} && m^{I}(y|\alpha_0,\sigma_0, M)
=\int_{0}^{\infty} \left( \int_{\Re^J}
f(y|\gamma,\sigma,M)p^{I}(\gamma|\sigma,\alpha_0,\sigma_0, M)
d\gamma \right) p^{I}(\sigma|\alpha_0,\sigma_0, M)
 d\sigma  \nonumber\\
&=&
\int_{0}^{\infty}
\left(
 \int_{\Re^J}
N_n(y|Z \gamma, \sigma^2 I_n) N_J(\gamma|\alpha_0 e,(\sigma^2+\sigma_0^2)W^{-1})d\gamma
\right)
p^{I}(\sigma|\alpha_0,\sigma_0, M)d\sigma,
\een

where $Z \equiv(  1_n \vdots X)$ and $\gamma^T \equiv (\alpha, \delta^T )$.

The inner integral in (\ref{mIP})  yields
$N_n(y|\alpha_0 Z e, \sigma^2 I_n+(\sigma^2 + \sigma_0^2)ZW^{-1}Z^T)$. 
This result can be shown directly or applying Lemma 3 of
\citet{MorenoEtAl:2003}.

Noticing that $Z e=1_n$, we now have to compute \ben \label{mIP1}
m^{I}(y|\alpha_0,\sigma_0, M)=\int _{0}^{\infty} N_n(y|\alpha_0
1_n, \sigma^2 I_n+ (\sigma^2 + \sigma_0^2)ZW^{-1}Z^T) \frac{2}{\pi
\sigma_0 (1+\frac{\sigma^2}{\sigma_0^2})} d\sigma. \een
An alternative expression for (\ref{mIP1}) is provided in the
supplementary material, together with an approximate evaluation
$\hat{m}^{I}(y|\alpha_0,\sigma_0, M)$.

Finally, the approximate BF based on the conditional intrinsic prior is computed as
  \be
   \widehat{BF}^{IP}_{M,M_0}(y|\alpha_0,\sigma_0)=\frac{\hat{m}^{I}(y|\alpha_0,\sigma_0, M)}{N_n(y| \alpha_0 1_n, \sigma_0^2 I_n)}.
\ee

\subsection{Bayes factor of a constrained model  relative to its encompassing model}
 In this section we deal with the computation of
 \ben
 \label{BFMcMe(c)}
BF^{IP}_{M_c,M_{e(c)}}(y|\alpha_0,\sigma_0)=
\frac{\prob^{I} \{ \theta \in \Theta_c|y,\alpha_0, \sigma_0, M_{e(c)}  \}}
{\prob^{I} \{ \theta \in \Theta_c|\alpha_0, \sigma_0, M_{e(c)}  \}}.
\een

Consider first the denominator of (\ref{BFMcMe(c)}). An analytical
evaluation is impossible; however the conditional intrinsic prior
as described in (\ref{CIPalphadeltasigma1}) and
(\ref{CIPalphadeltasigma2}) (with $M=M_{e(c)})$ lends itself to an
immediate estimate of the denominator, by iteratively sampling
values of $\sigma^2$ from (\ref{CIPalphadeltasigma2}) and then
sampling values of $(\alpha, \delta)$ from the conditional
distribution (\ref{CIPalphadeltasigma1}).
Let $( \alpha^{(t)},\delta^{(t)} )$, $t=1, \ldots, T$ be the
sampled values. Estimate the denominator of (\ref{BFMcMe(c)}) as
\be \widehat{\prob}^{I} \{\theta \in  \Theta_c|\alpha_0, \sigma_0,
M_{e(c)}  \} =\frac{\#  \{\delta^{(t)} \in \Theta_c \}}{T}.  \ee

Consider now the numerator of (\ref{BFMcMe(c)}). This involves
the  conditional intrinsic posterior distribution. For a generic
model $M$ described in (\ref{MsamplingModel}), letting
$\gamma^T=(\alpha, \delta^T )$ and expressing the prior in terms of the random
variable $\sigma^2$ instead of $\sigma$,  we have \be
\label{posteriorGammaSigmasq} p^{I}(\gamma,\sigma^2|y,\alpha_0,
\sigma_0, M) &\propto&
f(y|\gamma,\sigma,M)p^{I}(\gamma|\sigma,\alpha_0,\sigma_0, M)
p^{I}(\sigma^2|\alpha_0,\sigma_0, M) \nonumber \\
&\propto &
N_n(y|Z \gamma, \sigma^2 I_n)
N_J(\gamma|\alpha_0 e,(\sigma^2+\sigma_0^2)W^{-1})
InvBeta(\sigma^2|\frac{1}{2},\frac{1}{2}, \sigma_0^2).
\ee

Since the prior is not conjugate to the likelihood, the posterior
is not amenable to iterative direct
sampling as for the prior (\ref{CIPalphadeltasigma1}) and
(\ref{CIPalphadeltasigma2}).
However, we can resort to an MCMC implementation; see the supplementary material.

\section{Applications}
\label{sec:Applications}
\subsection{Simulation examples} 
In this subsection we evaluate the performance of our method through some simulation studies.
For comparison purposes we used the same setting presented in
 \cite{VanWesel:2011}.
The first example concerns ANOVA experiments for a few populations having  distinct
structures of group means, and homogeneous variances (homoschedasticity).   In the second
example  group variances are allowed to be different (heteroschedasticy).
We assume equal group sizes.

\vspace{0.4cm}
\textbf{Example 1}

The datasets are represented by 500 simulations, from each of five
populations  $\{1,2s, 2m, 2l, 3\}$, with five groups $j=1,\ldots,5$,  and separately for  two
group sizes $n_j=25$ and $n_j=50$. To each population there
corresponds a true generating model, according to the scheme
described in Table \ref{tab:ex1}.

{\renewcommand{\baselinestretch}{1}
\begin{table}[!h]
\centering \caption{\protect\small \textit{Example 1: group means
$\mu_j$, standard deviations $\sigma_j$, and true generating
model.}} \label{tab:ex1}
\begin{tabular}{|c|l|c|c|}
\hline
Pop & \multicolumn{1}{|c|}{$\mu_j$} & $\sigma_j$ & Model \\
\hline
1 & 0, 0, 0, 0, 0 & 1 & $M_0$ \\
2s & 0, 0.2, 0.4, 0.6, 0.8 & 1 & $M_2$ \\
2m & 0, 0.3, 0.6, 0.9, 1.2 & 1 & $M_2$ \\
2l & 0, 0.4, 0.8, 1.2, 1.6 & 1 & $M_2$ \\
3  & 2.23, 1.33, 3.23, 2.33, 3.23 & 1.55 & $M_3$ \\
 \hline
\end{tabular}
\end{table}
}
The competing models are
\be
M_0&:&\mu_1=\mu_2= \mu_3= \mu_4= \mu_5 \\
M_e&:&\mu_1,\mu_2, \mu_3, \mu_4, \mu_5 \\
M_2&:&\mu_1< \mu_2 <\mu_3< \mu_4 < \mu_5 \\
M_3&:&\mu_2< \mu_1 < \mu_4 < \{\mu_3 = \mu_5 \}. \\
\ee

We note that populations $\{2s, 2m, 2l\}$ correspond to
model $M_2$ with an increasing separation between adjacent means.
Our results are summarized in Table \ref{tab:ex1 prob}.

{\renewcommand{\baselinestretch}{1}
\begin{table}[!h]
\centering \caption{{\protect\small \textit{Example 1: percentage
over 500 simulations of  largest Bayes factors, and posterior model
probability medians ($PMP_{med}$) for the correct model  (in brackets the corresponding results of \citet{VanWesel:2011}). }}}
\label{tab:ex1 prob}%
\begin{tabular}{|c|c|c|c|c|c|c|}
\hline
Pop & $n_j$ &  $M_0$ & $M_2$ & $M_3$ & $M_e$& $PMP_{med}$ \\
\hline
\multirow{4}*{1}    & 25 &  \textbf{100} & 0 & 0 & 0 & 1.00  \\
      &    &  (89)        & (2) & (4) & (5) & (0.80) \\
      & 50 &  \textbf{100} & 0 & 0 & 0 & 1.00  \\
      &    &  (96)        & (1) & (2) & (1) & (0.93) \\
\hline
\multirow{4}*{2s}   & 25 &  82 & \textbf{7} & 5 & 6 & 0.02  \\
      &    &  (6)        & (79) & (13) & (2) & (0.75) \\
      & 50 &  43 & \textbf{49} & 1 & 7 & 0.42  \\
      &    &  (1)        & (92) & (7) & (0) & (0.93) \\
\hline
\multirow{4}*{2m} & 25 &  10 & \textbf{79} & 4 & 6 & 0.85  \\
   &    &  (0)        & (97) & (3) & (0) & (0.94) \\
      & 50 &  1 & \textbf{85} & 7 & 7 & 0.88  \\
      &    &  (0)        & (100) & (0) & (0) & (0.98) \\
\hline
\multirow{4}*{2l}    & 25 &  10 & \textbf{80} & 3 & 7 & 0.93  \\
    &    &  (0)        & (99) & (1) & (0) & (0.98) \\
      & 50 &  0  & \textbf{96} & 1 & 3 & 0.99  \\
      &    &  (0)        & (100) & (0) & (0) & (0.99) \\
\hline
\multirow{4}*{3}    & 25 &  4   & 0 & \textbf{96} & 0 & 0.93  \\
     &    &  (0)        & (0) & (96) & (4) & (0.94) \\
      & 50 &  0  & 0 & \textbf{100} & 0 & 1.00  \\
      &    &  (0)        & (0) & (96) & (4) & (0.96) \\
 \hline
\end{tabular}
\end{table}
}

For each
model we report  the percentage of times (out of the 500 simulations) in
which  the model obtained the highest  Bayes factor (\ref{BFIPMcM0});
 equivalently, it  scored the highest posterior
probability, because  we assume equal prior model probabilities.
Values in boldface correspond to the true model,
while values in  brackets are those computed by
\citet[Tables 5 and 6]{VanWesel:2011} -wherein the
null model is labeled as $M_1$, while the encompassing model is indicated as
$M_0$- using an empirical
Expected Posterior Prior (EEPP)  with an optimal  minimal training sample size equal to 2.
 The last column reports the median of the posterior
model probabilities (out of the 500 simulations) for the correct model
($PMP_{med}$).

Some broad features emerge from Table \ref{tab:ex1 prob}: when the null model $M_0$
holds, our method is able to capture it perfectly, slightly
improving on EEPP; a similar conclusion holds for population 3.
For population $2$ results differ depending on the level of
separation between consecutive means, and   on group
sample size. In particular, for population $2s$,  our method
favors $M_0$ when $n_j=25$, while it gives a 50\% chance to either
$M_0$ or $M_2$ when $n_j=50$. Although seemingly unsatisfactory, this result is indeed  quite sensible. To see why, consider for simplicity the comparison of two adjacent means, say $\mu_1$ and $\mu_2$. Letting $\Delta=\mu_2-\mu_1$, we have $\Delta=0$ under $M_0$ and $\Delta=0.2$ under $M_2$ (with  slight abuse of notation,  we use the same model symbols as in the example with five means). For given $\sigma_1$, $\sigma_2$, $n_1$ and $n_2$, and assuming $M_2$ is true,  can we confidently detect whether $\Delta>0$ (equivalently rule out $M_0)$?  One way to answer this query, from a frequentist perspective, is the following. A $95\%$ confidence interval (c.i.) for $\Delta$ is $D \pm 1.96*\mbox{sd}$, where $D=\bar{y}_2-\bar{y}_1$, and $\mbox{sd}$ is the standard deviation of $D$, namely $\mbox{sd}=\sqrt{\sigma^2_{1}/n_1+\sigma^2_{2}/n_2}$. To rule out, with high confidence,  $M_0$ in favor of $M_2$,  we would require  the c.i to be to entirely to the right of $\Delta=0$; equivalently $D>1.96*\mbox{sd}$. For a given  $\Delta>0$, we can compute $\prob \{ D>1.96*\mbox{sd} \}$, which is $\prob \{ Z>(1.96*\mbox{sd}-\Delta)/\mbox{sd}  \}$. This last expression is the \textit{power} of the experiment, i.e. the probability of excluding  $\Delta=0$ (with high confidence) when indeed $\Delta>0$ obtains. This probability is usually set at level 80\% in the social and health sciences, in order to determine an appropriate sample size for the experiment; see \citet[sects 20.2 and 20.3]{Gelm:Hill:2007} for several illustrations.
Table \ref{tab:ex1power} reports  the power for populations $\{2s, 2m, 2l\}$. It can be seen that the power is nowhere near 80\%. In particular for population $2s$ and $n_j=25$
it is only 10\%, and only rises to 17\% when $n_j=50$. The reason for this poor power is that the $\mbox{sd}$ associated to the estimator $D$ is $2\sigma/ \sqrt{n}$, with $\sigma_1=\sigma_2=\sigma$, and $n=n_1+n_2$. When $\sigma=1$ and $n_1=n_2=25$ we obtain $\mbox{sd}=0.28$, so that the two means are  only 0.20/0.28=0.71 units of $\mbox{sd}$ apart: models $M_0$ and $M_2$ ($2s$) are thus very poorly separated, and our method clearly reveals this, and opts for the more parsimonious choice $M_0$. Actually, given the relatively low powers for each of the three populations under  $M_2$, ranging between 0.10 and 0.52,  the performance of the objective Bayes approach seems remarkable at capturing the true generating model for population $2l$, and to a good extent also for $2m$. It would thus appear that, even when the power is only moderate (say of the order of 30\%), the Bayesian conclusion can be already quite firm, in terms of posterior probabilities for the correct model.

{\renewcommand{\baselinestretch}{1}
\begin{table}[!h]
\centering \caption{{\protect\small \textit{Example 1: power of
excluding $M_0$, when the true model is $M_2$. Calculations  are based on  pairwise comparisons and
a confidence interval at level 95\%. }}}
\begin{tabular}{|c|c|c|c|c|c|c|} \hline
\label{tab:ex1power}
Pop & $n_j$ &  $\Delta$ & $Power$  \\
\hline
\multirow{2}*{2s}   & 25 &  \multirow{2}*{0.2} & 0.10  \\
      & 50 &   & 0.17  \\
      \hline
\multirow{2}*{2m}   & 25 & \multirow{2}*{0.3} & 0.19  \\
      & 50 &   & 0.32  \\
\hline
\multirow{2}*{2l}   & 25 &  \multirow{2}*{0.4} & 0.30  \\
      & 50 &  & 0.52  \\
\hline
\end{tabular}
\end{table}
}

\vspace{0.4cm}
\textbf{Example 2}

In this example data were generated  either from the null model $M_0$, or from each of the three populations consistent with model $M_2$, as discussed in the previous example. However, to evaluate sensitivity to  model variances, each experiment was replicated under three distinct  heteroschedastic settings, characterized
by an increasing value of the ratio $F$
between the largest and
 smallest group variance; see Table \ref{tab:ex2}.
The results are summarized in Table \ref{tab:ex2 prob} according to the same format of Table \ref{tab:ex1 prob}.
The broad conclusion is that  our method is still capable of identifying the true generating model, with performances similar to those reported  by
\citet[Tables 8 and 9]{VanWesel:2011}. As already recalled in the discussion of Table \ref{tab:ex1 prob} based on power considerations,  the exception represented by population  $2s$ should  be of no concern.
We thus conclude that  our model selection procedure is  effective also under heteroskedasticity.

 {\renewcommand{\baselinestretch}{1}
 \begin{table}[!h]
 \centering \caption{{\protect\small \textit{Example 2: group means
 $\mu_j$ and true generating model (left panel);  standard
 deviations $\sigma_j$ and levels of violation of homogeneity
 assumption ($F$) (right panel). }}}
 \label{tab:ex2}%
 \begin{tabular}{|c|l|c|}
 \hline
 Pop & \multicolumn{1}{|c|}{$\mu_j$}&  \\
 \hline
 1 & 0, 0, 0, 0, 0          & $M_0$ \\
 2s & 0, 0.7, 1.4, 2.1, 2.8 & $M_2$ \\
 2m & 0, 1.1, 2.2, 3.3, 4.4 & $M_2$ \\
 2l & 0, 1.4, 2.8, 4.2, 5.6 & $M_2$  \\
  \hline
 \end{tabular}
 \quad
 \begin{tabular}{|c|c|}
 \hline
  $\sigma_j$ & $F$ \\
 \hline
  3, 3, 3, 3, 3 & 1\\
 1.4, 2.2, 3, 3.8, 4.6 & 11\\
  1, 2, 3, 4, 5 & 25\\
 \hline
 \end{tabular}

\end{table}
 }

{\renewcommand{\baselinestretch}{1}
\begin{table}[!h]
\centering \caption{{\protect\small \textit{Example 2: percentage
over 500 simulations of  largest Bayes factors, and posterior model
probability medians ($PMP_{med}$) for the correct model  (in brackets the corresponding results of \citet{VanWesel:2011}).}}}
\label{tab:ex2 prob}%
\begin{tabular}{|c|c|c|c|c|c||c|c|c|c|}
\hline
 & & \multicolumn{4}{|c||}{$n_j=25$} & \multicolumn{4}{|c|}{$n_j=50$} \\
\hline
Pop & F &  $M_0$ & $M_2$ & $M_1$& $PMP_{med}$ &  $M_0$ & $M_2$ & $M_1$& $PMP_{med}$  \\
\hline
\multirow{6}*{1}    & 1 &  \textbf{100}  & 0 & 0 & 1.00  & \textbf{100} & 0 & 0 & 1.00 \\
     &   & (92.4)  & (2.6) & (5) & (0.88)  & (97) & (1.4) & (1.6) & (0.96) \\
      & 11 &  \textbf{100} & 0 & 0 & 1.00  & \textbf{100} & 0 & 0 & 1.00 \\
      &   & (89.8)  & (4) & (6.2) & (0.89)  & (93.4) & (2.6) & (4) & (0.96) \\
      & 25 &  \textbf{100} & 0 & 0 & 1.00  & \textbf{100} & 0 & 0 & 1.00\\
      &   & (87.6)  & (6.2) & (6.2) & (0.88)  & (95.6) & (2) & (2.4) & (0.96) \\
\hline
\multirow{6}*{2s}    & 1 &  79  & \textbf{20} & 1 & 0.077 & 8 & \textbf{92} & 0 & 0.98  \\
      &   & (3.4)  & (95.2) & (1.4) & (0.92)  & (0) & (99.8) & (0.2) & (0.97) \\
      & 11 & 75 & \textbf{25} & 0 & 0.192  & 18 & \textbf{82} & 0 & 0.96 \\
      &   & (3.6)  & (95) & (1.4) & (0.93)  & (0.2) & (99) & (0.8) & (0.97) \\
      & 25 & 70 & \textbf{28} & 2 & 0.086  & 18 & \textbf{82} & 0 & 0.98 \\
       &   & (7)  & (90.6) & (2.4) & (0.92)  & (0.4) & (98) & (1.6) & (0.97) \\
\hline
\multirow{6}*{2m}    & 1 &  1  & \textbf{99} & 0 & 0.99 & 0 & \textbf{100} & 0 & 1.00 \\
      &   & (0)  & (100) & (0) & (0.98)  & (0) & (100) & (0) & (0.99) \\
      & 11 &  8 & \textbf{92} & 0 & 0.98  & 0 & \textbf{100} & 0 & 1.00\\
      &   & (0.2)  & (99.2) & (0.6) & (0.98)  & (0) & (99.8) & (0.2) & (0.99) \\
      & 25 &  10 & \textbf{90} & 0 & 0.98 & 0 & \textbf{100} & 0 & 1.00  \\
      &   & (0)  & (99.4) & (0.6) & (0.98)  & (0) & (99.8) & (0.2) & (0.99) \\
\hline
\multirow{6}*{2l}    & 1 &  0  & \textbf{100} & 0 & 1.00 & 0 & \textbf{100} & 0 & 1.00 \\
      &   & (0)  & (100) & (0) & (0.99)  & (0) & (100) & (0) & (0.99) \\
      & 11 &  0 & \textbf{100} & 0 & 0.99 & 0 & \textbf{100} & 0 & 1.00 \\
      &   & (0)  & (99.6) & (0.4) & (0.98)  & (0) & (100) & (0) & (0.99) \\
      & 25 &  0 & \textbf{100} & 0 & 0.99 & 0 & \textbf{100} & 0 & 1.00 \\
      &   & (0)  & (99.8) & (0.2) & (0.98)  & (0) & (99.8) & (0.2) & (0.99) \\
 \hline
\end{tabular}%
\end{table}
}

\subsection{Lucas' data}
\label{subsec:lucas}

The third example deals with real data.  The objective of this study
is to find out
what group members think about the competence of their leader;
see
\citet{Lucas:2003} for further details.
 It consists of five groups, each  having the same sizes
($n_j=30$) but different variances.
The five groups are:
randomly assigned male leader (1); randomly assigned female leader (2); male leader assigned
on ability (3); female leader assigned on ability (4);  institutionalized female leader (5).
The following four models represent substantive research interests
\be
M_a&:&\mu_2< \mu_1 < \mu_4 < \{\mu_3 = \mu_5 \} \\
M_b&:&\{\mu_1, \mu_3 \} > \{\mu_2,\mu_4,\mu_5\}  \\
M_0&:&\mu_1=\mu_2= \mu_3= \mu_4= \mu_5 \\
M_e&:&\mu_1,\mu_2, \mu_3, \mu_4, \mu_5,
\ee
with $M_0$ and $M_e$ denoting  the null, respectively  encompassing,  model.
The  results of our method are reported in Table \ref{tab:lucas}: there is an overwhelming evidence in favor
of model $M_a$, which actually corresponds Lucas' research hypothesis.

{\renewcommand{\baselinestretch}{1}
\begin{table}[!h]
\centering \caption{{\protect\small \textit{Lucas' data. Bayes
factors (BF) and Posterior Model Probabilities (PMP) for the
four competing models  (in brackets the corresponding results of \citet{VanWesel:2011}).}}}
 \label{tab:lucas}
\begin{tabular}{|c|c|c|}
\hline
Model & BF & PMP \\
\hline
$M_a$ & 5487.02  & 1.0  \\
      & (49.69)    & (0.98)  \\
$M_b$ & 0.00  & 0.00   \\
      & (0.01)   & (0.00) \\
$M_0$ & 0.00 &   0.00 \\
      & (0.01)  & (0.00) \\
$M_e$ &  1 & 0.00    \\
      & (1) & (0.02) \\
\hline
\end{tabular}%
\end{table}
}

\section{Discussion}
\label{sec:Discussion}

The  comparison of  models defined
through inequality and equality constraints on the parameter space  is of practical interest in several scientific areas.

In the frequentist setting,  the Akaike information criterion (AIC) is a  standard tool for model comparison: it
  contains two terms: one measuring fit and the other complexity. For  unconstrained models, the latter typically coincides with the number of  parameters. However the number of parameters in an inequality constrained model is the same as in the unconstrained one; as a consequence,  if the comparison has to be based on AIC-type criteria, suitable modifications of the complexity term are required.
 \citet{Anra:1999} proposed the order-restricted information (ORIC) for Gaussian ANOVA models. This criterion was extended in  \citet{Kuip:Hoij:Silv:2011} to the more general case wherein
population means may be restricted by a mix of linear equality and inequality constraints; the corresponding criterion has been named GORIC. Both criteria, which reduce to the usual AIC in the unconstrained case, encapsulate a component  of fit and  complexity, the latter being  strictly \textit{smaller} than the  number of parameters when an inequality constraint holds.
This formalizes our intuition that a constrained model is \lq \lq less complex\rq \rq{} than the unconstrained one.

In this paper we have presented an objective Bayesian approach for the comparison of models defined through inequality or equality constraints with special emphasis on normal ANOVA models. By comparing models in terms of the Bayes factor, a natural measure of fit and complexity is embodied in the marginal distribution for the observables.
When assessed with respect  to
alternative objective Bayes methodologies, notably the empirical expected posterior prior (EEPP),
our method is relatively inexpensive from a computational viewpoint, fully automatic, treats equality constraints exactly, and  produces comparable results in a variety of settings.
Remarkably our approach is able to identify the true generating model, relative to the null one, even in situations where frequentist-based power calculations would suggest a lack of effect-detectability.

In this work we have used a uniform prior on model space for the sake of simplicity and comparison with results obtained
using alternative methods. Other choices for priors on model space can be used in conjunction with our method;
see for instance \cite{Scot:Berg:2010} in the context of variable selection,  or  \cite{Carv:Scot} and \cite{Alt:Cons:LaRo:2013}
for graphical model determination.



\vspace{0.4cm}
\textbf{Acknowledgements.}
This work benefited from the comments of two reviewers. In particular, the critical points raised by one of them, led to the inclusion of section \ref{sec:Bayesian comparison}.
\bibliographystyle{ba}

\bibliography{Constrained}

\end{document}


\title{ Objective Bayesian Comparison of Constrained Analysis of Variance Models \\ \textit{Supplementary material}} 
\author{
Guido Consonni\\
Universit\`{a} Cattolica del Sacro Cuore, Milan, Italy\\
\url{guido.consonni@unicatt.it}
\and
Roberta Paroli \\
Universit\`{a} Cattolica del Sacro Cuore, Milan, Italy\\
\url{roberta.paroli@unicatt.it}
}

\date{  }

\maketitle

\section{Approximate evaluation of the integral (16)}
 We first provide an alternative expression for  the
integral 
(16) in the paper which is more suitable for
numerical evaluation.
Make the change of variable
 \be
 \eta=\frac{\sigma^2}{\sigma^2+\sigma_0^2},
 \ee
 so that
  $\sigma= \sqrt{\sigma_0^2\frac{\eta}{1-\eta}} $, with $0\leq \eta \leq 1$, and one can write the integrand in (16) 
  as a function of $\eta$.
  In particular the variance-covariance matrix of the normal density becomes
  $\frac{\sigma_0^2}{1-\eta}( \eta I_n+ Z W^{-1}Z^T)$.
  Additionally, since the prior
on $\sigma^2$ is  $InvBeta(\sigma^2|\frac{1}{2},\frac{1}{2},
\sigma_0^2)$,   it can be checked that the induced prior on $\eta$
is the  $Beta(1/2, 1/2)$ distribution. As a consequence we can
rewrite 
(16) as
  \ben
\label{mIP2} m^{I}(y|\alpha_0,\sigma_0, M)=\int _{0}^1
N_n(y|\alpha_0 1_n, \frac{\sigma_0^2}{1-\eta}( \eta I_n+ Z
W^{-1}Z^T)) Beta(\eta|\frac{1}{2},\frac{1}{2}) d\eta. \een
To evaluate (\ref{mIP2}) we use a method proposed by
\cite{Chib:1995},  and further extended in  \cite{Chib:Jeli:2001},  to approximate  a
marginal likelihood using an  MCMC algorithm.

For a given model $M$ with sampling density $f(y|\theta, M)$ for the data $y$,
and prior $p(\theta | M)$, the marginal likelihood $m(y|M)$
can be
written as: \be
m(y|M)=\frac{f(y|\theta, M)p(\theta|M)}{p(\theta|y, M)}. \ee
If $\theta^*$ is a high-density point in the support of the
posterior, such as the posterior mode or mean, or the maximum
likelihood estimate,
 an approximate expression for the  marginal likelihood is given by
\ben \label{chib} \ln \hat{m}(y|M)=\ln f(y|\theta^*, M) + \ln
p(\theta^*|M)- \ln \hat{p}(\theta^*|y, M). \een
The method first runs a Metropolis-Hastings (M-H) scheme  for $N$
iterations to obtain the posterior mode $\theta^*$, so that  the
first two addenda in the r.h.s. of (\ref{chib}) are computed. Next
the same M-H algorithm is run for further $2N$ iterations to
provide an estimate of the posterior density at the given value $\theta^*$ 
 \ben \label{chib1}
\hat{p}(\theta^*|y,M)=\frac{\sum_{i=1}^{N}\alpha(\theta^{(i)},\theta^*|y)
q(\theta^{(i)},\theta^*|y)}{\sum_{m=1}^{N}\alpha(\theta^*,\theta^{(m)}|y)},
\een
 where $\alpha(\theta,\theta^\prime|y)$ is the acceptance
probability of the proposed value and $q(\theta,\theta^\prime|y)$
is the proposal density for the transition from $\theta$ to
$\theta^\prime$, possibl depending on $y$. The sequence of the ${\theta^{(i)}}$ at the
numerator of (\ref{chib1}) is drawn  from the posterior
distribution, while that of the ${\theta^{(m)}}$ at the denominator is
drawn from the proposal $q(\theta,\theta^\prime|y)$.

%

When applied to our case,  we chose $q(\eta,\eta^\prime|y)$ equal
to the prior of $\eta$, so that  the acceptance probability is
\be
\alpha\left( \eta^{(k)}, \eta^{\ast } | y \right) =\min \left\{ 1;%
\frac{p^I(\eta^{(k)}|\gamma,y,\alpha_0,\sigma_0,M)p(\eta^{(k)})
}{p^I(\eta^{\ast}|\gamma,y,\alpha_0,\sigma_0,M)p(\eta^{\ast}) }
\right\}; \ee
 where the density
$p^I(\eta|\gamma,y,\alpha_0,\sigma_0,M)$ has the expression in
equation (\ref{fullCondEta}).

\section{Approximate evaluation of the BF in equation (17)}
We describe the MCMC procedure to approximate the numerator  of 
 equation (17). 
 The full conditionals of $(\gamma, \eta)$ are
\begin{itemize}
 \item \emph{Full conditional of $\gamma$} \be p^{I}(\gamma|\sigma,
y,\alpha_0, \sigma_0, M) &\propto&
f(y|\gamma,\sigma,M)p^{I}(\gamma|\sigma,\alpha_0,\sigma_0, M) \\
&\propto & N_n(y|Z \gamma, \sigma^2 I_n) N_J(\gamma|\alpha_0
e,(\sigma^2+\sigma_0^2)W^{-1}) \ee Standard calculations lead to
the conclusion that the full conditional of $\gamma$ is \be
\gamma|\sigma, y,\alpha_0, \sigma_0, M \sim
N_J(\gamma|\mu_{\gamma}(\sigma),\Sigma_{\gamma}(\sigma)), \ee where \be
\Sigma_{\gamma}(\sigma)=\left(\frac{W}{\sigma^2+\sigma_0^2}+\frac{Z^TZ}{\sigma^2}
\right)^{-1}, \, \mu_{\gamma}(\sigma)=\Sigma_{\gamma}(\sigma) \left(
\frac{W}{\sigma^2+\sigma_0^2} \alpha_0 e +\frac{Z^Ty}{\sigma^2}
\right). \ee

\item \textit{Full conditionals of $ \eta$} 

First consider the full conditional of $\sigma^2$
\be
p^{I}(\sigma^2|\gamma, y, \alpha_0,\sigma_0, M) &\propto & N_n(y|Z
\gamma, \sigma^2 I_n) N_J(\gamma|\alpha_0
e,(\sigma^2+\sigma_0^2)W^{-1})
InvBeta(\sigma^2|\frac{1}{2},\frac{1}{2}, \sigma_0^2) \\
&\propto& \left(\frac{1}{\sigma^2}\right)^{\frac{n+1}{2}} \exp
\left\{-\frac{C(\gamma)}{2\sigma^2} \right \}
\left(\frac{1}{\sigma^2+\sigma_0^2}\right)^{\frac{J+1}{2}} \exp
\left\{-\frac{D(\gamma)}{2(\sigma^2+\sigma_0^2)} \right \}, \ee
where \be C(\gamma)=(y-Z\gamma)^T(y-Z\gamma), \quad
D(\gamma)=(\gamma-\alpha_0e)^T W(\gamma-\alpha_0e). \ee 
Now make 
the change of variable \be \sigma^2 \mapsto
\eta=\frac{\sigma^2}{\sigma^2+\sigma_0^2}, \ee 
and taking into
account the Jacobian
$\frac{d \sigma^2}{d \eta}=\frac{1}{(1-\eta)^2}$, we obtain
\ben \label{fullCondEta} p^{I}(\eta|\gamma, y, \alpha_0,\sigma_0,
M) &\propto & 
\left(\frac{1}{\eta} \right)^{\frac{n+1}{2}}
(1-\eta)^{\frac{n+J-1}{2}} \nonumber  \\
& \times & \exp \{ -\frac{1}{2 \sigma_0^2}
[(1-\eta)(C(\gamma)+D(\gamma))+\frac{C(\gamma)}{\eta}] \}. \een
Since the normalizing constant of $p^{I}(\eta|\gamma,
y, \alpha_0,\sigma_0, M)$ is not analytically available,  one must resort to a Metropolis step to sample values of $\eta$.
\end{itemize}

In conclusion, the MCMC procedure to obtain a numerical
approximation of (17) 
can be summarized as follows. 
For $t=1, \ldots, T$ 
\bitem
\item [i)]
sample $\eta^{(t)}$ from $p^{I}(\eta|\gamma, y, \alpha_0,\sigma_0,
M_{e(c)})$ and compute
$(\sigma^2)^{(t)}=\sigma_0^2\frac{\eta^{(t)}}{1-\eta^{(t)}}$;
\item[ii)]
sample  $\gamma^{(t)}=( \alpha^{(t)},\delta^{(t)} )^T$,
 from
$N_J(\gamma|\mu_{\gamma}(\sigma^{(t)}),\Sigma_{\gamma}(\sigma^{(t)}))$;
\item[iii)]
compute \be \widehat{\prob}^{I} \{ \theta \in \Theta_c|y,\alpha_0,
\sigma_0, M_{e(c)}  \} =\frac{\#  \{\delta^{(t)} \in \Theta_c
\}}{T}. \ee  \eitem 
Finally obtain \be
 \label{BFMcMe(c)Estimate}
\widehat{BF}^{IP}_{M_c,M_{e(c)}}(y|\alpha_0,\sigma_0)=
\frac{\widehat{\prob}^{I} \{\theta \in  \Theta_c|y,\alpha_0,
\sigma_0, M_{e(c)}  \}} {\widehat{\prob}^{I} \{ \theta \in
\Theta_c|\alpha_0, \sigma_0, M_{e(c)}  \}}. \ee

\bibliographystyle{biometrika}

\bibliography{Constrained}